\documentclass[draftcls, onecolumn]{IEEEtran}
\IEEEoverridecommandlockouts
\usepackage{cite}
\usepackage{amsmath,amssymb,amsfonts,amsthm}
\usepackage{graphicx}
\usepackage{textcomp}
\usepackage{xcolor}
\usepackage{soul}
\usepackage{bm}
\usepackage{steinmetz}
\usepackage[noend]{algpseudocode}
\usepackage{lipsum}
\usepackage{algorithm}
\usepackage[caption=false,labelfont=sf,textfont=sf]{subfig}
\usepackage{multirow}
\usepackage{setspace}

\usepackage{tikz}
\usetikzlibrary{patterns}

\newtheorem{Lem}{Lemma}
\newtheorem{Thm}{Theorem}



\def\BibTeX{{\rm B\kern-.05em{\sc i\kern-.025em b}\kern-.08em
    T\kern-.1667em\lower.7ex\hbox{E}\kern-.125emX}}

\graphicspath{{./figure/}}

\begin{document}

\title{QoS-Aware Downlink Beamforming for Joint Transmission in Multi-Cell Networks \\ (with complementary materials)}

\author{\IEEEauthorblockN{Chen-Yen Lin},
\IEEEauthorblockA{\textit{MediaTek Inc.}, Hsinchu, Taiwan\\
\texttt{freddie.lin@mediatek.com}}\\
\and
\IEEEauthorblockN{Kuang-Hao (Stanley) Liu},
\IEEEauthorblockA{\textit{Institute of Communications Engineering},
\textit{National Tsing Hua University}, Hsinchu, Taiwan\\
\texttt{khliu@ee.nthu.edu.tw}}
}

\maketitle
\begin{abstract}
Multi-cell cooperation is an effective means to improve service quality to cellular users. Existing work primarily focuses on interference cancellation using all the degrees of freedom (DoF). This leads to low service quality for some users with poor channel quality to its serving base station. This work investigates the multi-cell beamforming design for simultaneously enhancing the downlink signal strength and mitigating interference. We first consider the ideal case when perfect channel state information (CSI) is available for determining the beamforming vectors and then extend to the case of imperfect CSI. For both cases, the beamforming optimization problems are non-convex. Assuming perfect CSI, we obtain the optimal joint transmit (JT) beamforming vectors based on the uplink-downlink duality. In the presence of unknown CSI errors, we use the semidefinite relaxation (SDR) with Bernstein-type inequality to derive the robust JT beamforming. Numerical results are presented to evaluate the performance of the proposed schemes.
\end{abstract}


\section{Introduction}

Multi-cell cooperation is a practical approach to enhance the service quality to subscribers in the cellular network. Through the coordination across multiple base stations (BSs) in different cells, a large virtual antenna array can be formed to increase the spatial degree of freedom (DoF), which may be used to enhance signal power by sending the same data to user equipment (UE), known as the joint transmission (JT). Alternatively, it can be used to mitigate interference by coordinated beamforming (CB). The increased DoF of multi-cell cooperation permits a flexible and robust design of multi-cell beamforming schemes that perform JT while simultaneously alleviating undesired interference, which is this work's objective.

In the literature, several works have explored multi-cell beamforming design with different objectives. As the single-cell counterpart, the summed throughput is a common design objective for multi-cell beamforming design~\cite{Yang2008}. Since it is challenging to solve the downlink (DL) throughput maximization problem directly, it is shown in~\cite{Yang2008} that reformulating the beamforming design as an uplink (UL) dual problem allows a more solvable form and the solution of the dual problem delivers the same summed throughput as the DL problem. Such a UL-DL dual relation has long been used to facilitate beamforming design in multiple-input multiple-output (MIMO) systems, such as the early work~\cite{Yu2006} for the single-cell MIMO system, and more recent work~\cite{Goettsch2022} for the cell-free MIMO system.  The UL-DL duality has also been used to minimize the total transmit power required to meet a prescribed quality-of-service (QoS) constraint, namely a target signal-to-interference-plus-noise ratio (SINR) in \cite{Dahrouj2010} assuming perfect CSI and in \cite{Wang2011, Shen2012} considering CSI errors. Again, they assume a single serving BS per UE~\cite{Dahrouj2010, Shen2011, Wang2011, Shen2012} and use all the DoF for interference mitigation. Another type of beamforming design considers fairness, as in~\cite{Gong2009} that jointly designs the transmit beamforming across multiple BSs to maximize the worst SINR among all the served UEs. Such a max-min formulation can also be tackled through the UL-DL duality. Unlike aforementioned works that use instantaneous CSI to design the beamforming, long-term CSI in terms of channel correlation matrices is adopted in~\cite{Gong2009} with the assumption of ideal CSI. Another popular method for solving the beamforming design problem is to linearize non-convex SINR expressions through the semi-definite relaxation (SDR)~\cite{Luo2010, Shen2011}, which provides a convex approximation to the non-convex optimization problem without closed-form solution. In contrast, the UL-DL duality is established based on the Lagrange duality that provides a closed-form solution for the optimal beamforming vectors. 

To fully leverage the benefit of multi-cell cooperation, this work explores the optimal JT beamforming at DL, where several BSs are grouped into a cluster and transmit DL data to the same UE simultaneously. This generalizes the existing work with the cluster size of one and can be implemented using the control signaling defined in the standard, such as 3GPP Release 16~\cite{TS38214}. This way, multiple DL data streams can be transmitted simultaneously to different UEs to increase spectral efficiency. We aim to minimize the total transmit power while satisfying a target SINR as in~\cite{Dahrouj2010, Shen2011, Wang2011, Shen2012} for CB. Unlike CB, our problem is more complicated because the transmit beamforming needs to maximize the summed signal power at the intended UE and suppress the interference to the other UEs, provided with the increased DoF via multi-cell cooperation. We show that the UL-DL duality can still be applied to obtain the optimal JT beamforming when global CSI is perfect. In practice, the backhaul links for CSI exchange have limited capacity, and thus the CSI exchanged across the cooperating BSs within the cluster is not perfect. The second part of this work addresses the optimal JT beamforming problem with imperfect CSI by extending~\cite{Wang2011} that considers the single-cell scenario. We demonstrate the significant gain of the optimal JT beamforming compared with the traditional non-JT beamforming and the zero-forcing  (ZF) based beamforming through simulation results. 

The remainder of this paper is organized as follows. Sec.~\ref{sec: model} explains the system model. Sec.~\ref{sec: proposed} presents the optimal JT beamforming design with perfect and imperfect CSI, respectively. Simulation results are shown in Sec.~\ref{sec: results}. Concluding remarks are given in Sec.~\ref{sec: conclusion}.
 
\section{System Model}\label{sec: model}

\begin{figure}[!t]
\centering
{
\includegraphics[width=0.75\linewidth]{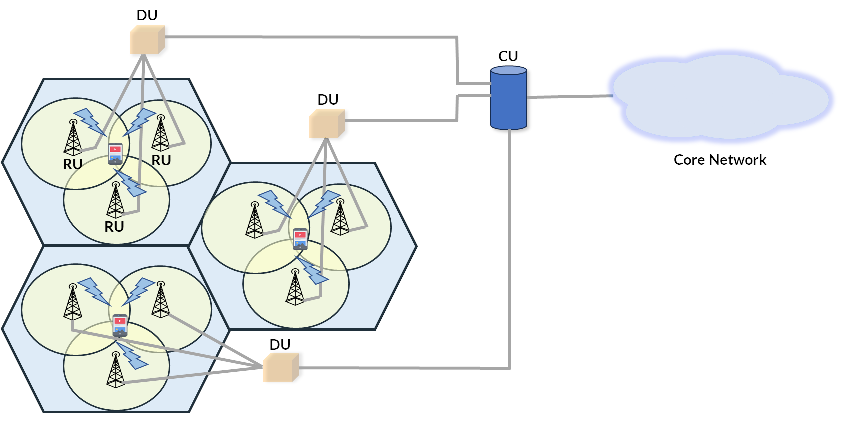}
}
\caption{Illustration of joint transmission in a multi-cell network. }
\label{fig: system-model}
\end{figure}

Consider a multi-cell network with multiple BSs deployed in a geographic area. Following the virtualized radio access network (vRAN) structure in 5G~\cite{Larsen2019}, a BS is split into different functional blocks, namely, the radio unit (RU) handling the RF transmission and reception, the distributed unit (DU) that carries out baseband signal processing, and the central unit (CU) performing the higher-layer protocols. Without loss of generality, one CU is deployed in the considered area, which is virtually partitioned into a set of cells denoted as $\mathcal{N}_c=\{\mathcal{C}_1,\cdots,\mathcal{C}_{N_c}\}$, each with $N_b$ RUs connected to one DU via fronthaul links. The DU collects CSI from the RUs within the same cell and designs the DL beamforming. In this work, the term RU and BS will be used interchangeably. Each BS has $N_t$ antennas, and each user equipment (UE) has a single antenna. 

For the $i$-th cell, denote $\mathcal{C}_i=\{\text{BS}_{ib}\}_{b=1}^{N_b}$ the set of BSs connected to the same DU for $i=1,\cdots,N_c$ and $\mathcal{U}_i=\{\text{UE}_{iu}\}_{u=1}^{|\mathcal{U}_i|}$ the set of UEs within the coverage area of the $i$-th cell. Considering the imperfect CSI, the true channel vector between $\text{BS}_{ib}$ and $\text{UE}_{iu}$ is modeled as
\begin{equation}\label{eq: imperfect channel}
\mathbf{h}_{ib,iu} =  \tilde{\mathbf{h}}_{ib,iu} + \mathbf{e}_{ib,iu} 
\end{equation}   
where $\tilde{\mathbf{h}}_{ib,iu} \in \mathbb{C}^{N_t}$ is the estimated channel vector, and $\mathbf{e}_{ib,iu}$ represents the CSI error, which is assumed to be a complex Gaussian random vector with zero mean and the covariance matrix $\mathbf{Q}_{ib,iu}=\epsilon_{ib,iu}^2 \mathbf{I}_{N_t}$ where $\epsilon_{ib,iu}^2$ is referred to as the error radius. Then we can express the CSI error vector as
\begin{equation}\label{eq: CSI error}
\mathbf{e}_{ib,iu} = \mathbf{Q}^{1/2}_{ib,iu} \mathbf{v}_{ib,iu}
\end{equation}
where $\mathbf{v}_{ib,iu}$ is a normalized complex random vector with zero mean and the covariance matrix $\mathbf{I}_{N_t}$.

For simplicity, we assume that a cluster consists of all the BSs within the same cell, i.e., BSs in each cluster are coordinated by one dedicated DU. In the JT mode, all the BSs in the cluster send the same signal to the intended UE. The received signal at $\text{UE}_{iu}$ is
\begin{align}
y_{iu} &= \sum_{b \in \mathcal{C}_i} \mathbf{h}_{ib,iu}^H \mathbf{w}_{ib,iu} s_{iu} + \sum_{b\in \mathcal{C}_i} \sum_{v \in \mathcal{U}_i \backslash u}  \mathbf{h}_{ib,iu}^H \mathbf{w}_{ib,iv} s_{iv} \nonumber \\
& \quad + \sum_{\mathcal{C}_j \in \mathcal{N}_c \backslash \mathcal{C}_i} \sum_{b'\in \mathcal{C}_j} \sum_{v'\in \mathcal{U}_j} \mathbf{h}_{jb',iu}^H \mathbf{w}_{jb',jv'} s_{jv'}+z_{iu}
\end{align}
where $s_{iu}$, $\mathbf{w}_{iu}\in \mathbb{C}^{N_t}$, and $z_{iu} \sim N(0,\sigma^2)$ represent the DL signal, the transmit beamforming vector,  and additive white Gaussian noise (AWGN) associated with $\text{UE}_{iu}$, respectively. The first term on the right-hand side is the collection of the intended signal for $\text{UE}_{iu}$, the second term is the intra-cell interference (Intra-CI), and the last term corresponds to the inter-cell interference (Inter-CI). Assuming the signal energy is normalized to one, the signal-to-interference-plus-noise ratio (SINR) of $\text{UE}_{iu}$ based on true CSI is given by
\begin{equation}\label{eq: sinr-downlink-perfect CSI}
\text{SINR}_{iu}(\mathbf{h}) = \frac{ \sum_{b \in \mathcal{C}_i} | \mathbf{h}_{ib,iu}^H \mathbf{w}_{ib,iu}|^2 }{I_\text{Intra} + I_\text{Inter} +  \sigma^2}
\end{equation}
where 
\begin{align}
I_{\text{Intra}} &= \sum_{b \in \mathcal{C}_i} \sum_{v \in \mathcal{U}_i \backslash u} | \mathbf{h}_{ib,iu}^H \mathbf{w}_{ib,iv}|^2\nonumber \\
I_{\text{Inter}} &= \sum_{\mathcal{C}_j \in \mathcal{N}_c \backslash \mathcal{C}_i} \sum_{b' \in \mathcal{C}_j} \sum_{v' \in \mathcal{U}_j} | \mathbf{h}_{jb',iu}^H \mathbf{w}_{jb',jv'}|^2
\end{align}


\section{Optimal Beamforming Design for JT}\label{sec: proposed}
Based on the considered system model, we formulate the optimal JT beamforming problem in this section, considering perfect and imperfect CSI, respectively.
\subsection{Problem Formulation}\label{sec: jt}

For $\text{UE}_{iu}$, the DU collects the CSI from all the BSs in $\mathcal{C}_i$ and designs the beamforming vectors jointly to minimize the total transmit power consumption for serving all the UEs while satisfying the SINR requirement of individual UE. Denote $\gamma_{iu}$ the SINR requirement of $\text{UE}_{iu}$, the JT beamforming design at DL can be formulated as
\begin{flalign}
(\text{JT-DL})  \quad  & \min_{\{\mathbf{w}_{ib,iu}\}} \sum_{\mathcal{C}_i \in \mathcal{N}_c} \sum_{b \in \mathcal{C}_i} \sum_{u \in \mathcal{U}_i}  \| \mathbf{w}_{ib,iu}  \|^2 & \nonumber \\
\text{s.t.} &~\text{SINR}_{iu}(\mathbf{h})\geq \gamma_{iu},~\forall~\mathcal{C}_i\in\mathcal{N}_c, u\in \mathcal{U}_i.& \label{eq: sinr constraint in JT-DL}
\end{flalign}
Due to the quadratic form of the beamforming vector $\mathbf{w}_{ib,iu}$ in $\text{SINR}_{iu}(\mathbf{h})$, (JT-DL) is a non-convex optimization problem. Besides, the beamforming vectors of different UEs are coupled, making it challenging to solve (JT-DL) directly. In this work, we employ the notion of \textit{uplink-downlink duality} to solve (JT-DL) efficiently.

\subsection{JT Beamforming with Perfect CSI}\label{sec: duality-JT}

\begin{table}[t!]
\centering
\caption{Dual relationship}
\label{Algo:nrjtbf}
	\begin{tabular}{|c || c c c|} 
	 \hline
	 Variable & Downlink & \quad & Uplink \\ [0.5ex] 
	 \hline\hline
	  Channel vector & $\mathbf{h}_{ib,iu}^H$ &  $\Leftrightarrow$ & $\mathbf{h}_{ib,iu}$ \\
      Beamforming vector & $\mathbf{w}_{ib,iu}$ & $\Leftrightarrow$ & $\hat{\mathbf{w}}_{ib,iu}$ \\ 
      Transmit Power & $\left\|\mathbf{w}_{ib,iu}\right\|^{2}$ &  $\Leftrightarrow$ & $\lambda_{iu} \sigma^{2}$ \\
	\hline
	\end{tabular}
\label{tab: duality}
\end{table}

The uplink-downlink duality refers to the equivalence of the uplink beamforming and the downlink beamforming. It implies the minimum transmit power at the BS required to achieve the DL SINR requirement is identical to that for the UL. Based on this principle, one can transform the downlink beamforming design problem to the uplink, which is computationally easier to solve. This idea has been adopted in~\cite{Dahrouj2010} for the non-JT case. This work extends the UL-DL duality to design the downlink beamforming for JT. Table~\ref{tab: duality} summarizes the UL-DL duality relation in our work.

The following theorem transforms (JT-DL) into the uplink counterpart. 

\begin{Thm}
The Lagrangian dual of (JT-DL) is the following uplink problem.
\begin{flalign}
(\text{JT-UL})   \quad  & \min_{\{\lambda_{iu}\}} \sum_{i=1}^{N_c} \sum_{u \in \mathcal{U}_i} \lambda_{iu}  \sigma^2 & \nonumber \\
\text{s.t.} &  \Lambda_{iu} \geq \gamma_{iu}& \label{eq: qos-ul} 
\end{flalign}
where $\lambda_{iu}$ is the dual variable of the $u$-th UE in the $i$-th cell and $\Lambda_{iu}$ is the UL SINR given by
\begin{equation}
\Lambda_{iu} = \frac{ \lambda_{iu} \sum_{b\in \mathcal{C}_i} | \hat{\mathbf{w}}^H_{ib,iu} \mathbf{h}_{ib,iu} |^2 }{I^{\mathrm{U}}_\mathrm{Intra}+ I^\mathrm{U}_\mathrm{Inter} + \sum_{b \in \mathcal{C}_i } \| \hat{\mathbf{w}}^H_{ib,iu} \|^2 }
\end{equation}
where $\hat{\mathbf{w}}_{ib,iu}$ is the UL receive beamforming vector given by
\begin{equation}\label{eq: optimal recv beamformer}
\hat{\mathbf{w}}_{ib,iu} = \left( \sum_{\mathcal{C}_j \in \mathcal{N}_c} \sum_{v \in \mathcal{U}_{j}} \lambda_{jv} \sigma^2 \mathbf{h}_{ib,jv} \mathbf{h}_{ib,jv}^H +\sigma^2 \mathbf{I}_{N_t} \right)^{-1} \mathbf{h}_{ib,iu}.
\end{equation}
In addition, $I^\mathrm{U}_\mathrm{Intra}$ and $I^\mathrm{U}_\mathrm{Inter}$ are given by
\begin{subequations}
\begin{align}
I^\mathrm{U}_\mathrm{Intra} &= \sum_{b \in \mathcal{C}_i} \sum_{v\in \mathcal{U}_i \backslash u} \lambda_{iv}  |\hat{\mathbf{w}}^H_{ib,iu} \mathbf{h}_{ib,iv}|^2, \\
I^\mathrm{U}_\mathrm{Inter} &= \sum_{\mathcal{C}_j  \in \mathcal{N}_c \backslash \mathcal{C}_i} \sum_{b'\in \mathcal{C}_{j}} \sum_{v' \in \mathcal{U}_{j}} \lambda_{jv'}\sigma^2 |\hat{\mathbf{w}}^H_{ib,iu}\mathbf{h}_{ib,jv'}|^2.
\end{align}
\end{subequations}
\end{Thm}
\begin{proof}
The proof is based on the fact that the non-convex SINR constraints in (JT-DL) can be transformed into the second-order cone constraints following~\cite{Wiesel2006}. Accordingly, the duality gap is zero, and strong duality holds for (JT-DL). This suggests that (JT-DL) can be solved by its Lagrangian dual problem. To this end, we first define the Lagrangian of (JT-DL) as given by
\begin{align}\label{eq: Lagrangian_v1}
&L(\mathbf{w}_{ib,iu} \lambda_{iu}) = \sum_{\mathcal{C}_i \in \mathcal{N}_c} \sum_{b\in \mathcal{C}_i} \sum_{u \in \mathcal{U}_i } \mathbf{w}_{ib,iu}^H \mathbf{w}_{ib,iu} - \sum_{\mathcal{C}_i \in \mathcal{N}_c} \sum_{u \in \mathcal{U}_i} \lambda_{iu} \nonumber \\
& \quad \Bigl[ \gamma_{iu}^{-1} \sum_{b\in \mathcal{C}_i} | \mathbf{h}_{ib,iu}^H \mathbf{w}_{ib,iu}|^2 - \sum_{b\in \mathcal{C}_i}  \sum_{v \in \mathcal{U}_i \backslash u}  | \mathbf{h}_{ib,iu}^H \mathbf{w}_{ib,iv} |^2 \nonumber \\
&  \quad - \sum_{\mathcal{C}_j \in \mathcal{N}_c \backslash \mathcal{C}_i} \sum_{b'\in \mathcal{C}_j} \sum_{v' \in \mathcal{U}_{j}} |\mathbf{h}^H_{jb',iu} \mathbf{w}_{jb',jv'}|^2 - \sigma^2 \Bigr]
\end{align}
where $\lambda_{iu}>0$ is the Lagrange multiplier associated with the SINR constraint of $\text{UE}_{iu}$. The dual function is
\begin{equation}
g(\lambda_{iu}) = \min_{\mathbf{w}_{ib,iu}}  L(\mathbf{w}_{ib,iu}), \forall~i,b,u.
\end{equation}
Then the Lagrangian dual of (JT-DL), namely, the maximum of $g(\lambda_{iu})$  is
\begin{flalign}
(\text{JT-DL Dual})  \quad  & \max_{\lambda_{iu}} \sum_{\mathcal{C}_i \in \mathcal{N}_c}  \sum_{u \in \mathcal{U}_i} \lambda_{iu} \sigma^2 & \nonumber \\
\text{s.t} & \sum_{b\in \mathcal{C}_i} \boldsymbol{\Theta}_{ib} \succeq (1+\gamma_{iu}^{-1}) \lambda_{iu} \mathbf{h}_{ib,iu}  \mathbf{h}_{ib,iu}^H, & \label{eq: uplink sinr constraint}
\end{flalign}
where 
\begin{equation}\label{eq: Theta_ib}
\bm{\Theta}_{ib} = \sum_{\mathcal{C}_i \in \mathcal{N}_c} \sum_{u \in \mathcal{U}_i} \lambda_{iu} \mathbf{h}_{ib,iu}+\mathbf{I}_{N_t}
\end{equation}
To show (JT-DL Dual) is equivalent to (JT-DL), we first find the optimal receive beamforming vector $\mathbf{w}_{ib,iu}$ that maximizes $\Lambda_{iu}$, which is the minimum-mean-square-error (MMSE) receive beamforming as given by \eqref{eq: optimal recv beamformer}. By substituting \eqref{eq: optimal recv beamformer} into \eqref{eq: qos-ul}, the uplink SINR constraint \eqref{eq: qos-ul} is identical to \eqref{eq: uplink sinr constraint}. Hence, (Dual-JT-DL) is the same as (JT-UL) except that the minimization is replaced by maximization, and the inequality constraints are reversed. In addition, both problems have the optimal solution when the SINR constraints are satisfied with equality. Therefore, (JT-DL Dual) and (JT-UL) have the same solution. 
\end{proof}
The optimal receive beamforming in \eqref{eq: optimal recv beamformer} depends on the dual variable $\lambda_{iu}$, which must satisfy the first-order necessary condition of the optimality. To find $\lambda_{iu}$, we first combine all the terms related to $\mathbf{w}_{ib,iu}$ in \eqref{eq: Lagrangian_v1}, which yields
\begin{align}\label{eq: Lagrangian_v2}
&L(\mathbf{w}_{ib,iu},\lambda_{iu}) = \sum_{\mathcal{C}_i \in  \mathcal{N}_c} \sum_{u \in \mathcal{U}_i } \lambda_{iu} \sigma^2 + \sum_{\mathcal{C}_i \in  \mathcal{N}_c} \sum_{b \in \mathcal{C}_i}  \sum_{u \in \mathcal{U}_i} \mathbf{w}_{ib,iu}^H \nonumber \\
&\Biggl[ \mathbf{I}_{N_t} - (1+ \gamma_{iu}^{-1}) \lambda_{iu} \mathbf{h}_{ib,iu} \mathbf{h}^H_{ib,iu} + \sum_{\mathcal{C}_j \in  \mathcal{N}_c} \sum_{v \in \mathcal{U}_j} \lambda_{jv} \mathbf{h}_{ib,jv}  \mathbf{h}_{ib,jv}^H \Biggr] \nonumber \\
&\times \mathbf{w}_{ib,iu}.
\end{align}
Taking the gradient of $L(\mathbf{w}_{ib,iu},\lambda_{iu})$ with respective to $\mathbf{w}^H_{ib,iu}$ and setting the result to zero, we obtain the necessary condition for optimal $\lambda_{i u}$ as
\begin{align}
\lambda_{i u}=\frac{1}{\left(1+\gamma_{iu}^{-1}\right) \sum_{b \in \mathcal{C}_i} \mathbf{h}_{ib,iu}^{H} \bm{\Theta}_{ib}^{-1} \mathbf{h}_{ib,iu}},
\label{fun:lambda}
\end{align}
which is a function of $\bm{\Theta}_{ib}$ given in \eqref{eq: Theta_ib} that depends on $\lambda_{iu}$. This suggests that one can iteratively optimize $\lambda_{iu}$ and $\hat{\mathbf{w}}_{ib,iu}$ until convergence. 

Based on the obtained receive beamformer $\hat{\mathbf{w}}_{ib,iu}$ and the UL-DL duality, the optimal DL JT beamforming can be derived from the following theorem.
\begin{Thm} 
\label{thm:scaledversionw}
The optimal DL JT beamforming ${\mathbf{w}}_{ib,iu}$ is a scaled version of the UL JT beamforming $\hat{\mathbf{w}}_{ib,iu}$ as given by
\begin{align}\label{thm2}
\mathbf{w}_{ib,ij}=\sqrt{\delta_{ib,u}}{\hat{\mathbf{w}}}_{ib,iu}.
\end{align}
The scaling variable $\delta_{ib,u}$ can be obtained as
\begin{align}\label{fun:delta}
\bm{\Delta}=\mathbf{F}^{-1}\mathbf{1}\mathbf{\sigma}^2
\end{align}
where $\bm{\Delta}=[\bm{\delta}_1,\bm{\delta}_2,\dots,\bm{\delta}_{N_c}]^{T} \in \mathbb{R}^{N_c N_b|\mathcal{U}_i|\times 1}$ with $\bm{\delta}_i=[\delta_{ib,u}]_{b \in \mathcal{C}_i,u\in \mathcal{U}_i}^{T}$ is an $N_b|\mathcal{U}_i|\times 1$ vector, $\mathbf{1}$ is an all-one vector of dimension $N_c|\mathcal{U}_i| \times 1$ and $\mathbf{F}$ is an $N_c |\mathcal{U}_i| \times N_c N_b |\mathcal{U}_j|$ matrix as given by
\begin{equation}\label{fun:F}
\mathbf{F}=
\begin{bmatrix}
\mathbf{F}_{1,11} & \dots & \mathbf{F}_{1,1  |\mathcal{C}_1|} & \mathbf{F}_{1,21} & \dots & \mathbf{F}_{1,N_c  |\mathcal{C}_{1}|}\\
\mathbf{F}_{2,11} & \dots & \mathbf{F}_{2,1 |\mathcal{C}_2|}& \mathbf{F}_{2,21}& \dots & \mathbf{F}_{2,N_c |\mathcal{C}_2|}\\
\vdots&\ddots&\vdots&\vdots&\ddots&\vdots\\
\mathbf{F}_{N_c,11} & \dots & \mathbf{F}_{N_c,  1|\mathcal{C}_{N_c}|}& \mathbf{F}_{{N_c},21}& \dots & \mathbf{F}_{{N_c, N_c  |\mathcal{C}_{N_c}|}}
\end{bmatrix}
\end{equation}
where the ($u$,$v$)-th entry of each $|\mathcal{U}_i| \times |\mathcal{U}_j|$  sub-matrix $\mathbf{F}_{i,jb}$ is defined as
\begin{equation}\label{fun:subF}
\mathbf{F}_{i,jb}(u,v)=\begin{cases}
\gamma_{iu}^{-1} |{\hat{\mathbf{h}}}_{ib,iu}^H{\hat{\mathbf{w}}}_{ib,iu}|^2, &~\mathrm{if}~{j}={i}~\mathrm{and}~v=u,\\
- |\tilde{\mathbf{h}}_{ib,iu}^H{\hat{\mathbf{w}}}_{ib,iv}|^2, &~\mathrm{if}~{j}={i}~\mathrm{and}~v\neq u,\\
-|{\tilde{\mathbf{h}}}_{mb,iu}^H{\hat{\mathbf{w}}}_{jb,jv}|^2, &~\mathrm{if}~{j}\neq {i}.
\end{cases}
\end{equation}
\end{Thm}
\begin{proof}
The proof can be found in Appendix.
\end{proof}

\begin{algorithm}[ht!]
\setstretch{1}
  \caption{Non-Robust JT Beamforming (NR-JTBF)}
  \label{Algo:nrjtbf}
  \begin{algorithmic}[1]
   \renewcommand{\algorithmicrequire}{\textbf{Input:}}
   \renewcommand{\algorithmicensure}{\textbf{Output:}}
   \Require CSI $\mathbf{h}_{ib,iu}$, $\forall~\mathcal{C}_i \in \mathcal{N}_c$, $b \in \mathcal{C}_i$, $u \in \mathcal{U}_i$.   
    \State Initialize $\lambda_{iu}=1$, $\forall~i, u$.
    \Repeat 
    \State Compute $\bm{\Theta}_{ib}$ using \eqref{eq: Theta_ib}.
    \State Compute $\lambda_{iu}$ using \eqref{fun:lambda}.
    \State Compute $\hat{\mathbf{w}}_{ib,iu}$ using \eqref{eq: optimal recv beamformer}.
	\State Compute the scaling factor $\delta_{i,bu}$ from \eqref{fun:delta}.
    \State Update $\mathbf{w}_{ib,iu}$ using \eqref{thm2}.
    \Until 
	the predefined stopping criterion is met. 
    \Ensure \\
    Optimal downlink JT beamformers $\mathbf{w}_{ib,iu},\forall~i,b,u$.
  \end{algorithmic}
\end{algorithm}
A complete algorithm for iteratively obtaining the optimal DL JT beamforming vectors $\mathbf{w}_{ib,iu}$ is summarized in Algorithm~\ref{Algo:nrjtbf}, which is referred to as non-robust JT beamforming (NR-JTBF) in this work. The convergence of Algorithm~\ref{Algo:nrjtbf} to the global optimal solution is guaranteed following the similar arguments in~\cite{Dahrouj2010}, which is omitted here due to space limit.

 

\subsection{JT Beamforming with Imperfect CSI}\label{sec: passive-beamforming-design}

Algorithm~\ref{Algo:nrjtbf} uses the accurate CSI as the input. In practice, CSI acquired by each BS contains unknown estimation errors and thus is imperfect. Consequently, the beamforming vectors derived from imperfect CSI may not achieve the required SINR. This motivates us to revisit (JT-DL) and design robust beamforming for JT that minimizes the total transmit power while ensuring the target SINR with a certain probability. The resultant problem formulation based on estimated CSI is given below.

%

\begin{flalign}
(\text{Robust JT-DL})  \quad  & \min_{\mathbf{w}_{ib,iu}} \sum_{\mathcal{C}_i \in \mathcal{N}_c} \sum_{b\in \mathcal{C}_i} \sum_{u \in \mathcal{U}_i}  \| \mathbf{w}_{ib,iu}\|^2 & \nonumber \\
\text{s.t} &~\Pr[ \text{SINR}_{iu}(\tilde{\mathbf{h}}) \geq \gamma_{iu} ] \geq 1-\rho_{iu}\label{eq: probabilistic sinr constraint}
\end{flalign}
where $\rho_{iu} \in (0,1)$. By imposing constraint \eqref{eq: probabilistic sinr constraint}, $\text{UE}_{iu}$ is guaranteed to satisfy the target SINR with the probability no less than $1-\rho_{iu}$. (Robust JT-DL) is non-convex since the beamforming vectors $\mathbf{w}_{ib,iu}$ are of quadratic forms and constraint \eqref{eq: probabilistic sinr constraint} is a probability function without closed-form. Unfortunately, UL-DL duality can only handle deterministic constraints. To remedy this, we use the SDR technique to convert (Robust JT-DL) into a convex problem.

Firstly, we linearize the objective function of (Robust JT-DL) by introducing the rank-one positive semidefinite (PSD) matrix $\mathbf{W}_{ib,iu} = \mathbf{w}_{ib,iu} \mathbf{w}^H_{ib,iu}$. As a result, the quadratic term $\|\mathbf{w}_{ib,iu}\|^2$ is replaced by $\text{Tr}(\mathbf{W}_{ib,iu} )$, which is a linear operation. Meanwhile,  constraint \eqref{eq: probabilistic sinr constraint} can be rewritten as
\begin{align}\label{MinTP3_C_rewrite3}
&\Pr\Bigg[\gamma_{iu}^{-1} \sum_{b \in \mathcal{C}_i} (\tilde{\mathbf{h}}_{ib,iu} + \mathbf{e}_{ib,iu})^H \mathbf{W}_{ib,iu}  (\tilde{\mathbf{h}}_{ib,iu} + \mathbf{e}_{ib,iu}) \nonumber \\
&- \sum_{b \in \mathcal{C}_i} \sum_{v \in \mathcal{U}_i \backslash u} (\tilde{\mathbf{h}}_{ib,iu} + \mathbf{e}_{ib,iu})^H \mathbf{W}_{ib,iv}  (\tilde{\mathbf{h}}_{ib,iu} + \mathbf{e}_{ib,iu}) \nonumber \\
&- \sum_{\mathcal{C}_j \in \mathcal{N}_c \backslash \mathcal{C}_i} \sum_{b' \in \mathcal{C}_j} \sum_{u' \in \mathcal{U}_j} (\tilde{\mathbf{h}}_{jb',iu} + \mathbf{e}_{jb',iu})^H \mathbf{W}_{jb',just}  \nonumber \\
& \quad (\tilde{\mathbf{h}}_{jb',iu} + \mathbf{e}_{jb',iu}) \geq \sigma^2 \Biggr] \geq 1-\rho_{iu},
\end{align}
which does not have a closed form and thus is difficult to handle. To simplify \eqref{MinTP3_C_rewrite3} into a more tractable form, we substitute \eqref{eq: CSI error} into \eqref{MinTP3_C_rewrite3} and define the following notations.
\begin{align}
& \bm{\Omega}_{ib,iu}\triangleq\mathbf{Q}_{ib,iu}^{1/2}(\gamma_{iu}^{-1}\mathbf{W}_{ib,iu}-\sum_{v \in \mathcal{U}_i \backslash u}\mathbf{W}_{ib,iv})\mathbf{Q}_{ib,iu}^{1/2}, \nonumber \\
& \bm{\Omega}_{jb',iu}\triangleq -\mathbf{Q}_{jb',iu}^{1/2}\sum_{u' \in \mathcal{U}_j} \mathbf{W}_{jb',ju'} \mathbf{Q}_{jb',iu}^{1/2}, \nonumber\\
&\bm{\omega}_{ib,iu} \triangleq \mathbf{Q}_{ib,iu}^{1/2}(\gamma_{iu}^{-1}\mathbf{W}_{ib,iu}-\sum_{v \in \mathcal{U}_i \backslash u}\mathbf{W}_{ib,iv}){\tilde{\mathbf{h}}}_{ib,iu}, \nonumber\\
& \bm{\omega}_{jb',iu} \triangleq -\mathbf{W}_{jb',iu}^{1/2} \sum_{u' \in \mathcal{U}_j} \mathbf{W}_{jb',ju'}  \tilde{\mathbf{h}}_{jb',iu},\nonumber \\
&d_{ib,iu} \triangleq \tilde{\mathbf{h}}_{ib,iu}^H (\gamma_{iu}^{-1} \mathbf{W}_{ib,iu}-\sum_{v \in \mathcal{U}_i} \mathbf{W}_{ib,iv}) \tilde{\mathbf{h}}_{ib,iv}, \nonumber \\
& d_{jb',iu}\triangleq -\tilde{\mathbf{h}}_{jb',iu}^H \sum_{u' \in \mathcal{U}_j} \mathbf{W}_{jb',ju'}) \tilde{\mathbf{h}}_{jb',iu}.
\label{fun:notation}
\end{align}
Then \eqref{MinTP3_C_rewrite3} can be expressed as
\begin{align}
&\Pr \Bigg[ \sum_{\mathcal{C}_l \in \mathcal{N}_c} \sum_{b \in \mathcal{C}_l} \Big(\mathbf{v}_{lb,iu}^H \bm{\Omega}_{lb,iu} \mathbf{v}_{lb,iu}+\bm{\omega}_{lb,iu}^H \mathbf{v}_{lb,iu} \nonumber \\
&+\mathbf{v}_{lb,iu}^H \bm{\omega}_{lb,iu}+d_{lb,iu}\Big) \geq\sigma^2 \Biggr ]\geq 1-\rho_{iu},~\forall \; \mathcal{C}_i \in \mathcal{N}_c, u\in \mathcal{C}_i.
\label{fun:MinTP5_Crewrite4}
\end{align}
Note that $\bm{\omega}_{lb,iu}^H \mathbf{v}_{lb,iu}+\mathbf{v}_{l\beta,iu}^H \bm{\omega}_{lb,iu} = 2\mathrm{Re}(\mathbf{v}_{lb,iu}^H \bm{\omega}_{lb,iu})$.
Hence, \eqref{fun:MinTP5_Crewrite4} can be represented as
\begin{align}
&\Pr \Biggl[ \sum_{\mathcal{C}_l \in \mathcal{N}_c} \sum_{b \in \mathcal{C}_l} \Big(\mathbf{v}_{lb,iu}^H \bm{\Omega}_{lb,iu} \mathbf{v}_{lb,iu} + 2\mathrm{Re}(\mathbf{v}_{lb,iu}^H \bm{\omega}_{lb,iu} \bigr) \nonumber \\
&\geq \sigma^2 -\sum_{\mathcal{C}_l\in \mathcal{N}_c} \sum_{\beta \in \mathcal{C}_l} d_{lb,iu} \Biggr] \geq 1-\rho_{iu},~\forall \; \mathcal{C}_i \in \mathcal{N}_c, u\in \mathcal{C}_i.
\label{fun:constraintrewrite}
\end{align}
Let $K = N_c N_b$. Then we can rewrite \eqref{fun:constraintrewrite} as
\begin{align}\label{fun:constraintrewrite2}
&\Pr \Biggl[ \sum_{k=1}^K \Big( \hat{\mathbf{v}}_{k,iu}^H \hat{\bm{\Omega}}_{k,iu} \hat{\mathbf{v}}_{k,iu} + 2\mathrm{Re}(\hat{\mathbf{v}}_{k,iu}^H \hat{\bm{\omega}}_{k,iu} \bigr) \nonumber \\
& \geq \sigma^2 - \sum_{k=1}^K \hat{d}_{k,iu} \Biggr] \geq 1-\rho_{iu}.
\end{align}
where $\hat{\mathbf{v}}_{k,iu} = \mathbf{v}_{lb,iu}$ for $k=(l-1)N_b+b$. Likewise, $\hat{\bm{\Omega}}_{k,iu}= \bm{\Omega}_{lb,iu} \in \mathbb{H}^{N_t}$, $\hat{\bm{\omega}}_{k,iu} = \bm{\omega}_{lb,iu} \in \mathbb{C}^{N_t}$ and $\hat{d}_{k,iu} = d_{lb,iu}$. Here $\mathbb{H}^{N_t}$ represents the $N_t$-dimensional Hermitian matrix. Although  \eqref{fun:constraintrewrite2} is much simpler than \eqref{MinTP3_C_rewrite3}, it remains a probabilistic inequality without a closed form, and thus a certain approximation is necessary. Since \eqref{fun:constraintrewrite2} contains the quadratic form of $\hat{\mathbf{v}}_{k,iu}^H$, we are motivated to apply the Bernstein-type inequality, which bounds the probability of the quadratic form of complex random variables deviating from its mean. To this end, define 
\begin{equation}\label{eq: g_kiu}
g_{k,iu} =  \hat{\mathbf{v}}_{k,iu}^H \hat{\bm{\Omega}}_{k,iu} \hat{\mathbf{v}}_{k,iu}+2\mathrm{Re}\left( \hat{\bm{\omega}}_{k,iu}^H \hat{\mathbf{v}}_{k,iu}\right),
\end{equation}
$\forall~k=1,\ldots,K$, $\mathcal{C}_i \in \mathcal{N}_c$, and $u \in \mathcal{U}_i$ . Then the following Lemma is introduced.
\begin{Lem}
For any $\tau_{iu}>0$, 
\begin{align}
&\Pr\Biggl[ \sum_{k=1}^K g_{k,iu} \geq \sum_{k=1}^K \mathrm{Tr}(\hat{\bm{\Omega}}_{k,iu})-\sqrt{2\tau_{iu}} \Bigl(\sum_{k=1}^K \big( \| \hat{\bm{\Omega}}_{k,iu} \|^2 \nonumber \\
& \quad +2 \| \hat{\bm{\omega}}_{k,iu} \|^2\bigr) \Bigr)^{1/2}-\tau_{iu} \lambda^{-}(\hat{\bm{\Omega}}) \Biggr] \geq 1-e^{-\tau_{iu}}.
\label{fun:btypeiq}
\end{align}
where $\lambda^{-}(\hat{\bm{\Omega}})=\max\lbrace{\lambda_{\max}(-\hat{\bm{\Omega}}_{1,iu}),\ldots,\lambda_{\max}(-\hat{\bm{\Omega}}_{K,iu}),0}\rbrace$ with $\lambda_{\max}(-\hat{\bm{\Omega}}_{k,iu})$ denoting the maximum eigenvalue of the matrix $(-\hat{\bm{\Omega}}_{k,iu})$. 
\label{lemma:bernstein}
\end{Lem}
\begin{proof}
The proof is motivated from~\cite{Xu2017}, which considers robust hybrid beamforming without multi-cell cooperation. The idea is to find a conservative convex approximation to the probabilistic constraint \eqref{fun:constraintrewrite2} with the aid of the Bernstein-type inequality~\cite{Bechar2009}. For our problem, a Bernstein-type inequality can be established as follows.

Let $S=K\cdot 2N_t$ for $K=N_cN_b$, $\bm{\alpha}=[\alpha_1,\ldots,\alpha_S]^T\in\mathbb{R}^S$ and $\bm{\beta}=[\beta_1,\ldots,\beta_S]^T\in\mathbb{R}^S$, where $\mathbb{R}^S$ denotes the $S$-dimensional real vector. Besides, $\bm{\nu}=[\nu_1,\ldots,\nu_S]^T\in\mathbb{R}^S$ be a random vector where $\nu_1,\ldots,\nu_S$ are i.i.d. real random variables following the standard normal distribution $\mathcal{N}(0,1)$. Define
\begin{align}
G=\sum_{s=1}^{S}{\Big(\alpha_s \nu_s^2+2\beta_s \nu_s\Big)}.
\label{fun:G}
\end{align}
Then, given $\tau>0$, the following result holds.
\begin{align}
&\Pr\left[ G\geq\sum_{s=1}^S \alpha_s-2\sqrt\tau\sqrt{\sum_{s=1}^S \Big({\alpha_s}^2+2{\beta_s}^2\Big)}-2\tau \alpha^-\right] \nonumber \\
 \geq & 1-e^{-\tau},
\label{fun:bto}
\end{align}
where $\alpha^{-}{\triangleq} \max\lbrace{\max \lbrace -\alpha_s \mid1\leq s \leq S}\rbrace,0\rbrace$. \eqref{fun:bto} is a real-valued constraint but can be applied to handle complex-valued variables. This is achieved with the aid of the following variables.
\begin{subequations}
\begin{align}
& \bar{\bm{\Omega}}_{k,iu}= \frac{1}{2}
\begin{bmatrix} 
\text{Re} ( \hat{\bm{\Omega}}_{k,iu} ) & -\text{Im} (\hat{\bm{\Omega}}_{k,iu} ) \\ \text{Im}(\hat{\bm{\Omega}}_{k,iu})  & \text{Re}( \hat{\bm{\Omega}}_{k,iu} ) \end{bmatrix} \in\mathbb{S}^{2N_t}, \label{eq: Omega bar}\\
& \bar{\bm{\omega}}_{k,iu} = \frac{1}{\sqrt2}
\begin{bmatrix} \text{Re} (\hat{\bm{\omega}}_{k,iu} ) \\ \text{Im} ( \hat{\bm{\omega}}_{k,iu} ) \end{bmatrix} \in\mathbb{R}^{2N_t}, \label{eq: omega bar}\\
& \bar{\mathbf{v}}_{k,iu} = \sqrt2 \begin{bmatrix} \text{Re} (\mathbf{v}_{k,iu} ) \\ \text{Im} (\mathbf{v}_{k,iu}) \end{bmatrix} \sim\mathcal{CN}\left(\mathbf{0},\mathbf{I}_{2N_t} \right), \label{eq: v bar}
\end{align}
\end{subequations}
for all $k \in \{1,\ldots,K\}, \mathcal{C}_i \in \mathcal{N}_c$, and $u \in \mathcal{U}_i$. Denote $\bar{v}_{k,iu}[l]$  and $\bar{\omega}_{k,iu}[l]$ the $l$th element of $\bar{\mathbf{v}}_{k,iu}$ and $\bar{\bm{\omega}}_{k,iu}$, respectively. Here, $\mathbb{S}$ denotes the real symmetric matrix. Using \eqref{eq: Omega bar}-\eqref{eq: v bar},  and decomposing $\bar{\bm{\Omega}}_{k,iu}$ as $\bar{\bm{\Omega}}_{k,iu}=\mathbf{U}_{k,iu} \bm{\Lambda}_{k,iu} \mathbf{U}_{k,iu}^T$ through eigenvalue decomposition, where $\mathbf{U}_{k,iu}$ is an orthogonal matrix and $\bm{\Lambda}_{k,iu}=\text{diag}(\lambda_{k,iu}[l])$ is a diagonal matrix with $\lambda_{k,iu}[l]$ denoting the $l$-th eigenvalue of $\bar{\bm{\Omega}}_{k,iu}$, \eqref{eq: g_kiu} can be rewritten as
\begin{align}
g_{k,iu} &=\bar{\mathbf{v}}_{k,iu}^T \bar{\bm{\Omega}}_{k,iu} \bar{\mathbf{v}}_{k,iu}+2 \bar{\bm{\omega}}_{k,iu}^T \bar{\mathbf{v}}_{k,iu} \nonumber \\
&\overset{(\text{i})}{=} \tilde{\mathbf{v}}_{k,iu}^T \bm{\Lambda}_{l,iu} \tilde{\mathbf{v}}_{k,iu}^T +2 \tilde{\bm{\omega}}_{k,iu}^T \tilde{\mathbf{v}}_{k,iu}  \nonumber\\
&\overset{(\text{ii})}{=} \sum_{l=1}^{2N_t} \Big(\lambda_{k,iu}[l] \tilde{v}^2_{k,iu}[l] +2 \tilde{\omega}_{k,iu}[l] \tilde{v}_{k,iu}[l]\Big),
\label{fun:gi1}
\end{align}
where (i) is obtained by defining $\tilde{\mathbf{v}}_{k,iu} = \mathbf{U}_{k,iu}^T \bar{\mathbf{v}}_{k,iu}$ and $\tilde{\bm{\omega}}_{k,iu} = \mathbf{U}_{k,iu}^T \bar{\bm{\omega}}_{k,iu}$, and in (ii), $\tilde{v}_{k,iu}[l]$ and $\tilde{\omega}_{k,iu}[l]$ is the $l$-th element of $\tilde{\mathbf{v}}_{k,iu}$ and $\tilde{\bm{\omega}}_{k,iu}$, respectively. One can see that \eqref{fun:gi1} follows the same form as \eqref{fun:G}.  Similar to \eqref{fun:G}, denote $G'$ the sum of $g_{k,iu},~\forall~k=1,\ldots,K$, i.e., 
\begin{align}
G' &= \sum_{k=1}^K g_{k,iu} \nonumber \\
&= \sum_{k=1}^K \sum_{l=1}^{2N_t} \Big(\lambda_{k,iu}[l] \tilde{v}^2_{k,iu}[l] +2 \tilde{\omega}_{k,iu}[l] \tilde{v}_{k,iu}[l]\Big).
\end{align}
According to~\eqref{fun:bto}, we obtain the following inequality. 
\begin{align}
&\Pr\Bigl[ G'\geq \sum_{k=1}^K \sum_{l=1}^{2N_t} \lambda_{k,iu}[l]- 2\sqrt{\tau_{iu}} \nonumber \\
& \sqrt{\sum_{k=1}^K \sum_{l=1}^{2N_t} \lambda^2_{k,iu}[l]+2\tilde{\omega}_{k,iu}^2[l] }-2\tau_{iu} \tilde{\alpha}^-\Bigr] \geq  1-e^{-\tau_{iu}},
\label{fun:bto-1}
\end{align}
where 
\begin{equation*}
\tilde{\alpha}^{-}{\triangleq} \max \{ \max_l \{-\lambda_{1,iu}[l] \},\ldots, \max \{ \max_l \{-\lambda_{K,iu}[l], 0 \},
\end{equation*}
for all $\mathcal{C}_i \in \mathcal{N}_c$, $u \in \mathcal{U}_i$. \eqref{fun:bto-1} can be written more compactly by using the property of matrix trace, e.g., $\sum_{l=1}^{2N_t} \lambda_{k,iu}[l] = \text{Tr}(\bar{\bm{\Omega}}_{k,iu})$ and also, $\bar{\bm{\Omega}}_{k,iu}$ has the same eigenvalues as $\hat{\bm{\Omega}}_{k,iu}$. Hence, \eqref{fun:bto-1}  can be rewritten as
\begin{align}
&\Pr\Bigl[ G'\geq \sum_{k=1}^K \text{Tr}(\hat{\bm{\Omega}}_{k,iu})- 2\sqrt{\tau_{iu}} \nonumber \\
& \sqrt{\sum_{k=1}^K \frac{1}{2} \| \hat{\bm{\Omega}}_{k,iu} \|^2 + \| \hat{\bm{\omega}}_{k,iu} \|^2 }-\tau_{iu} \lambda^-(\hat{\bm{\Omega}}) \Bigr] \geq  1-e^{-\tau_{iu}},
\label{fun:bto-2}
\end{align}
which is obtained by noting that
\begin{subequations}
\begin{align}
&\sum_{l=1}^{2N_t} \lambda^2_{k,iu}[l] = \| \bar{\bm{\Omega}}_{k,iu} \|^2 = \frac{1}{2} \| \hat{\bm{\Omega}}_{k,iu} \|^2, \\
&2\sum_{l=1}^{2N_t}\tilde{\omega}_{k,iu}^2[l] = 2 \| \tilde{\bm{\omega}}_{k,iu} \|^2 = 2 \| \bar{\bm{\omega}}_{k,iu} \|^2 = \| \hat{\bm{\omega}}_{k,iu} \|^2 \\
&\tilde{\alpha}^- =  \max\{ \lambda_{\max}(-\bar{\bm{\Omega}}_{1,iu}), \cdots, \lambda_{\max}(-\bar{\bm{\Omega}}_{K,iu}), 0\} \nonumber  \\
&\quad \; = \frac{1}{2}\max\{ \lambda_{\max}(-\hat{\bm{\Omega}}_{1,iu}), \cdots, \lambda_{\max}(-\hat{\bm{\Omega}}_{K,iu}), 0\} \nonumber \\
&\quad \; \triangleq \frac{1}{2} \lambda^-(\hat{\bm{\Omega}}).
\end{align}
\end{subequations}
It is easy to check that \eqref{fun:bto-2} is identical to \eqref{fun:btypeiq} that completes the proof.
\end{proof}

By letting $\rho_{iu} = e^{-\tau_{iu}}$ and comparing \eqref{fun:btypeiq} with \eqref{fun:constraintrewrite2}, we can see that the following inequality serves as a sufficient condition to satisfy  \eqref{fun:constraintrewrite2}.
\begin{align}
&\sum_{k=1}^K \mathrm{Tr}(\hat{\bm{\Omega}}_{k,iu})-\sqrt{2\tau_{iu}} \Bigl(\sum_{k=1}^K \big( \| \hat{\bm{\Omega}}_{k,iu} \|^2 +2 \| \hat{\bm{\omega}}_{k,iu} \|^2\bigr) \Bigr)^{1/2} \nonumber \\
&-\tau_{iu} \lambda^{-}(\hat{\bm{\Omega}}) \geq \sigma^2 - \sum_{k=1}^K \hat{d}_{k,iu}, \quad \forall \; \mathcal{C}_i \in \mathcal{N}_c, u\in \mathcal{U}_i.
\label{fun:iq2}
\end{align}
Recall $K = N_c N_b$ and $k=(l-1)N_b+b$, \eqref{fun:iq2} can be rewritten as
\begin{align}
&\sum_{\mathcal{C}_l \in \mathcal{N}_c} \sum_{b \in \mathcal{C}_l}  \Bigl(\mathrm{Tr}(\bm{\Omega}_{lb,iu})+d_{lb,iu} \Bigr)- \tau_{iu} \lambda^{-}(\bm{\Omega})-  \sigma^2 \geq \nonumber \\
& \sqrt{2\tau_{iu}} \Bigl(\sum_{\mathcal{C}_l \in \mathcal{N}_c} \sum_{b \in \mathcal{C}_l}  \|\mathbf{C}_{lb,iu}\|^2 \Bigr)^{1/2}.
\label{fun:iq4}
\end{align}
where $\mathbf{C}_{lb,iu}=[\bm{\Omega}_{lb,iu}~\sqrt{2}\bm{\omega}_{lb,iu}]$. Since $\lambda^{-}(\bm{\Omega})$ in \eqref{fun:iq4} is non-convex, we introduce the slack variable $x_{iu} \geq 0$ such that \eqref{fun:iq4} is equivalent to
\begin{align}
&\sqrt{\sum_{\mathcal{C}_j \in \mathcal{N}_c} \sum_{b \in \mathcal{C}_j} \Vert  \mathbf{C}_{jb,iu}\Vert^2}\leq \frac{1}{\sqrt{2\tau_{iu}}}  \Big[\sum_{\mathcal{C}_j \in \mathcal{N}_c} \sum_{b \in \mathcal{C}_j}\Big(\text{Tr}  (\bm{\Omega}) +d_{jb,iu}\Big)\nonumber\\
&~~~~-\tau_{iu}x_{iu}-\sigma^2 \Bigr],\nonumber
\end{align}
along with $x_{iu} \mathbf{I}_{N_t}+\mathbf{Q}_{jb,iu} \succeq \mathbf{0}$. As a result, we obtain a convex formulation for (Robust JT-DL) as follows.
\begin{flalign}
&(\text{Robust JT-DL}^2)&\\
&  \quad   \min_{\mathbf{W}_{ib,iu}, x_{iu} \geq 0} \sum_{\mathcal{C}_i \in \mathcal{N}_c} \sum_{b\in \mathcal{C}_i} \sum_{u \in \mathcal{U}_i}   \text{Tr}(\mathbf{W}_{ib,iu}) & \nonumber \\
\text{s.t} & \quad \sqrt{\sum_{\mathcal{C}_j \in \mathcal{N}_c} \sum_{b \in \mathcal{C}_j} \Vert  \mathbf{C}_{jb,iu}\Vert^2}\leq \frac{1}{\sqrt{2\tau_{iu}}}  &\nonumber \\
 & \quad \times \Big[\sum_{\mathcal{C}_j \in \mathcal{N}_c} \sum_{b \in \mathcal{C}_j}\Big(\text{Tr}  (\bm{\Omega}) +d_{jb,iu}\Big)-\tau_{iu}x_{iu}-\sigma^2 \Bigr], &\\
 & \quad  x_{iu} \mathbf{I}_{N_t}+\mathbf{Q}_{lb,iu} \succeq \mathbf{0}, ~\forall~i, b, u.&
\end{flalign}
With the aid of SDR formulation and the Bernsteirn-type inequality, (Robust JT-DL$^2$) is convex that can be solved numerically. The obtained solution $\mathbf{W}_{ib,iu}$, which is referred to as robust-JT beamforming (R-JTBF), is not necessarily a rank-one matrix, in which case the Gaussian randomization technique can be used to find an approximated solution\cite{Luo2010}.

\section{Simulation Results}\label{sec: results}

The performance of the proposed NR-JTBF and R-JTBF are evaluated by simulating the three-cell scenario in Fig.~\ref{fig: system-model}. In each cell, a UE is randomly distributed and 10,000 channel realizations are generated for each UE location. The channel vector between $\text{BS}_{ib}$ and $\text{UE}_{iu}$ is modeled as~$\mathbf{h}_{ib,iu} = \sqrt{\beta_{ib,iu}} \psi_{ib,iu} \varphi_{ib,iu}( \tilde{\mathbf{h}}_{ib,iu} + \mathbf{e}_{ib,iu} )$, where $\beta_{ib,iu}$ and $\psi_{ib,iu}$ capture the distance-dependent pathloss and lognormal shadowing, respectively, and $\varphi_{ib,iu}$ reflects the transmit-receive antenna gain.  Besides, $\tilde{\mathbf{h}}_{ib,iu} \sim \mathcal{CN}(\mathbf{0}, \mathbf{I}_{N_t})$ is the Rayleigh fading coefficient. Table~\ref{tab:ex} lists the simulation parameters. 

\begin{table}[h]\label{tab:ex}
\centering
\caption{Simulation parameters}
\begin{tabular}[phbt!]{|c|c|}
\hline \bf{Parameter} & \bf{Value} \\
\hline Number of cells $N_c$ & 3 \\
\hline Number of BSs per cell $N_b$ & 3 \\
\hline Number of UEs per cell & 1 \\
\hline Number of BS antennas $N_t$ & 4 \\
\hline Inter-cell distance & 500 m\\
\hline BS maximum power & 24 dBm \\
\hline Antenna gain $\varphi_{ib,iu}$ & 5 dBi \\
\hline System bandwidth & 10 MHz \\
\hline Noise power spectral density & -174 dBm/Hz \\
\hline Pathloss over a distance of $d$ km &  $145.4+37.5 \log_{10}d$ \\
\hline CSI error radius $\epsilon$ & $\sqrt{0.1}$ \\
\hline SINR requirement $\gamma_{iu}$ & 12 dB \\
\hline SINR violation probability $\rho_{iu}$ & 0.1 \\
\hline
\end{tabular}
\end{table}

\begin{figure}[!t]
\centering
{
\includegraphics[width=1\linewidth]{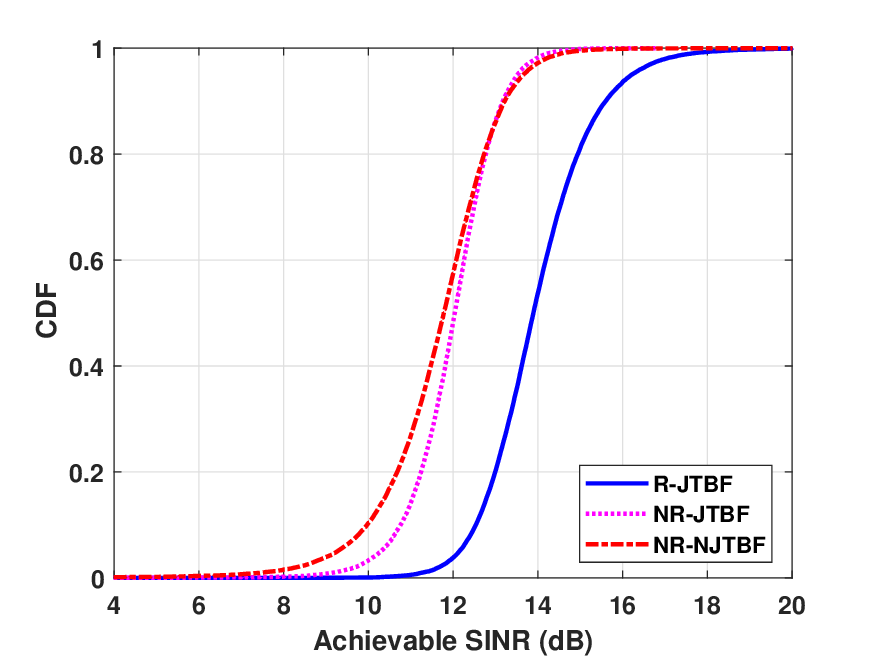}
}
\caption{The CDF of the achievable SINR.}
\label{fig: sinr cdf}
\end{figure}

We first show the distribution of the achievable SINR in terms of the cumulative distribution function (CDF) in Fig.~\ref{fig: sinr cdf}. Here, the traditional multi-user MIMO beamforming based on imperfect CSI without JT referred to as non-robust non-JT beamforming (NR-NJTBF), is included for comparison. NR-NJTBF is a special case of the proposed NR-JTBF with one BS per cell, i.e., $N_b=1$. One can see that R-JTBF significantly improves the achievable SINR over NR-JTBF and NR-NJTBF. The probability of violating the SINR target of $12$ dB is only $0.04$, which is much less than the prescribed probability of 0.1. As to NT-JTBF and NR-NJTBF, the likelihood of SINR target violation is $49\%$ and $61\%$, respectively. The result justifies the importance of robust beamforming for practical systems with imperfect CSI.

\begin{figure}[!t]
\centering
\includegraphics[width=1\linewidth]{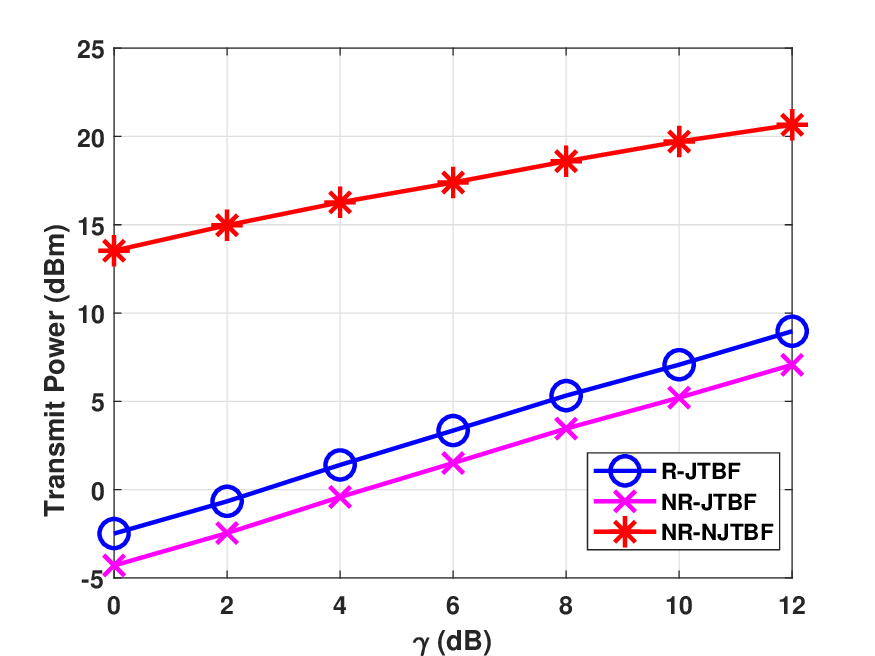}
\caption{Transmit power vs. SINR target $\gamma$.}
\label{fig: power vs sinr target}
\end{figure}

Fig.~\ref{fig: power vs sinr target} plots the transmit power used by different beamforming schemes for satisfying the SINR target $\gamma$, which is assumed to be identical for all UEs. As $\gamma$ increases, a higher transmit power is required to meet the SINR target as expected. Given a specific SINR target, the proposed R-JTBF consumes extra power of 2 dBm compared with the proposed NR-JTBF. The additional power consumption of R-JTBF can be regarded as the cost for compensating the SINR loss due to imperfect CSI and the same price is maintained even for a more stringent SINR target. Comparing with NR-NJTBF, both the proposed R-JTBF and NR-JTBF dramatically reduce the required transmit power. For example, R-JTBF consumes 12 dBm less power than NR-NJTBF when $\gamma=12$ dB. The power-saving gain attributes to JT, which uses the antennas across different BSs and, thus, it achieves a better beamforming gain than the beamforming scheme without JT.

\begin{figure}[!t]
\centering
{
\includegraphics[width=1\linewidth]{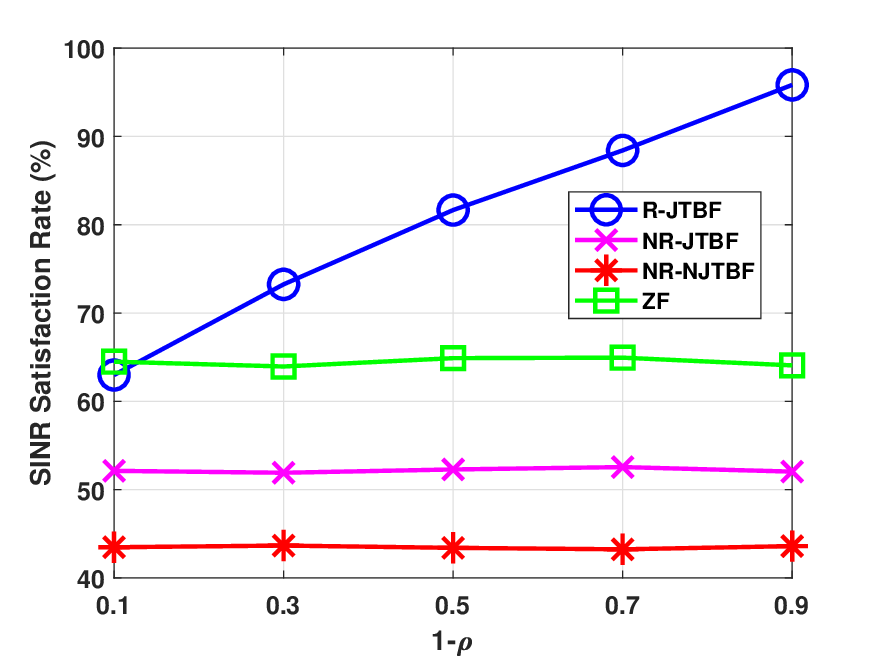}
}
\caption{SINR satisfaction rate vs. predefined SINR satisfaction probability.}
\label{fig: sinr satisfaction rate vs 1-rho}
\end{figure}

We further examine the performance of R-JTBF under varied QoS requirements in terms of the designated SINR satisfaction probability $1-\rho$. Here, the commonly considered multi-user beamforming, namely zero-forcing (ZF), is included in the comparison. The ZF beamforming vectors, as given by $\mathbf{w}_{ib,iu}^{\text{ZF}} = \mathbf{h}_{ib,iu}( \mathbf{h}_{ib,iu}^H \mathbf{h}_{ib,iu})^{-1}$ for $\text{UE}_{iu}$ served by $\text{BS}_{ib}$, aim to cancel all the interference, including intra-cell and inter-cell interference by jointly designing the beamforming using all the antennas in the network. This can be implemented at the CU, which collects global CSI. Fig.~\ref{fig: sinr satisfaction rate vs 1-rho} shows the SINR satisfactory rate achieved by different schemes versus the designated SINR satisfactory probability. Except ZF without considering a specific QoS requirement, the same SINR target of 6 dB is applied to the rest schemes. In addition, the SINR satisfaction probability $(1-\rho)$ is varied from 0.1 to 0.9 for R-JTBF. The SINR satisfaction rate of R-JTBF consistently exceeds the designated satisfaction probability and is higher than the rest schemes. For the three non-robust beamforming schemes, the SINR satisfaction rate of ZF, NR-JTBF, and NR-NJTBF is about $63\%$, $51\%$, and $42\%$, respectively. It is worth pointing out that ZF outperforms NR-JTBF and NR-NJTBF because it utilizes all the available antennas in the network to serve each user, while the rest two use the antennas within the same cell only. Thus, the gain of ZF is the direct consequence of using $N_c$ times more antennas than that of NR-JTBF and $N_c N_b$ times more antennas than that of NR-NJTBF with a single-BS operation.

\begin{figure}[!t]
\centering
{
\includegraphics[width=1\linewidth]{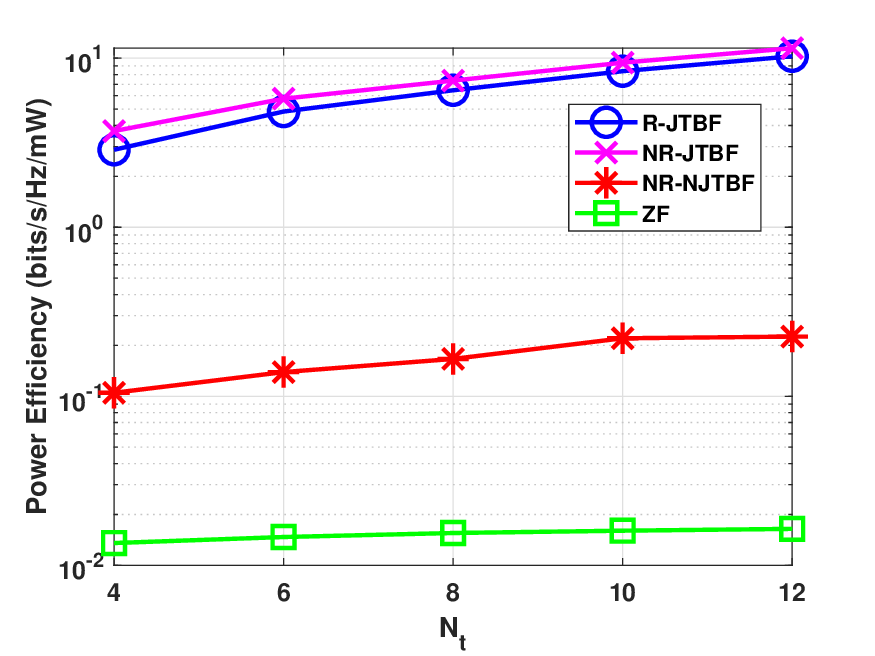}
}
\caption{Power efficiency vs. number of antennas per BS.}
\label{fig: power efficiency vs Nt}
\end{figure}

Finally, we investigate how much spectral efficiency can be improved per unit of power. To this end, we define the power efficiency as the ratio between the spectral efficiency evaluated as $\sum_{\mathcal{C}_i  \in \mathcal{N}_c} \sum_{u \in \mathcal{U}_i}  \log_2(1+\text{SINR}_{iu})$ and the total transmit power. Fig.~\ref{fig: power efficiency vs Nt} illustrates the power efficiency as a function of the antenna per BS $N_t$. Increasing $N_t$ leads to improved SINR due to a higher spatial DoF. Also, the required power to meet the SINR target is reduced by using a larger antenna array at the BS. As a result, each unit of energy can be used more efficiently, as seen from the increasing trend in Fig.~\ref{fig: power efficiency vs Nt}. Both JT beamforming schemes, namely R-JTBF and NR-JTBF, have a much higher power efficiency than NR-NJTBF and ZF. The energy efficiency of R-JTBF is slightly worst than that of NR-JTBF because of a higher transmit power paid to achieve the desired SINR subject to CSI error as seen in Fig.~\ref{fig: power vs sinr target}. On the other hand, ZF has the worst energy efficiency because it uses all the available power at the BS. As a result, ZF is not energy efficient despite a certain spectral efficiency increase as $N_t$ increases.

\section{Conclusion}\label{sec: conclusion}
This work extends the previous work on the multi-cell coordinated beamforming design to the JT mode. It is shown that when perfect CSI is not available, the robust JT beamforming scheme significantly improves the achievable SINR, thereby increasing the SINR satisfaction rate over the non-robust JT beamforming scheme and ZF-based beamforming scheme. The improvement of the robust JT beamforming scheme comes at the cost of slightly increased transmit power compared to the non-robust JT beamforming scheme. Still, the cost is nearly negligible as the number of antennas per BS increases. On the other hand, the non-robust non-JT beamforming scheme incurs a dramatic increase in the total transmit power as the target SINR increases. Hence, its power efficiency is much lower than the two JT beamforming schemes. The results of this work demonstrate that signal enhancement and interference cancellation can be achieved simultaneously through multi-cell cooperation that yields promising gains even without perfect CSI. 

\appendix
We prove in this appendix that the optimal DL JT beamformer $\mathbf{w}_{ib,iu}$ is a scaled version of the optimal dual UL JT receive beamformer $\hat{\mathbf{w}}_{ib,iu}$ given in \eqref{eq: optimal recv beamformer}. we start by taking the gradient of the Lagrangian in \eqref{eq: Lagrangian_v2} with respect to $\mathbf{w}_{ib,iu}^H$ and setting the result to zero that yields
\begin{align}\label{eq: result of diff}
&\Biggl[ \mathbf{I}_{N_t} - (1+ \gamma_{iu}^{-1}) \lambda_{iu} \mathbf{h}_{ib,iu} \mathbf{h}^H_{ib,iu} + \sum_{\mathcal{C}_j \in  \mathcal{N}_c} \sum_{v \in \mathcal{U}_j} \lambda_{jv} \mathbf{h}_{ib,jv}  \mathbf{h}_{ib,jv}^H \Biggr] \nonumber \\
&\times \mathbf{w}_{ib,iu} = 0.
\end{align}
By rearranging \eqref{eq: result of diff}, we get 
\begin{align}
\Biggl[ \mathbf{I}_{N_t} +  \sum_{\mathcal{C}_j \in  \mathcal{N}_c} \sum_{v \in \mathcal{U}_j} \lambda_{jv} & \mathbf{h}_{ib,jv}  \mathbf{h}_{ib,jv}^H \Biggr] \mathbf{w}_{ib,iu} \nonumber \\
&= \left[ (1+ \gamma_{iu}^{-1}) \lambda_{iu} \mathbf{h}_{ib,iu} \mathbf{h}^H_{ib,iu} \right] \mathbf{w}_{ib,iu}.
\end{align}
Multiply both sides by $\sigma^2$ and after some arrangement, we have
\begin{align}\label{eq: scaling equation}
\mathbf{w}_{ib,iu} =&  \underbrace{ \Bigl(  \sum_{\mathcal{C}_j \in  \mathcal{N}_c} \sum_{v \in \mathcal{U}_j} \lambda_{jv} \sigma^2 \mathbf{h}_{ib,jv}  \mathbf{h}_{ib,jv}^H + \sigma^2 \mathbf{I}_{N_t} \Bigr)^{-1} \mathbf{h}_{ib,iu} }_{\hat{\mathbf{w}}_{ib,iu} } \nonumber \\
& \times \Big(  [\sigma^2 (1+\gamma_{iu})^{-1}) \lambda_{iu} \mathbf{h}_{ib,iu}^H ] \mathbf{w}_{ib,iu} \Bigr),
\end{align}
which shows that $\mathbf{w}_{ib,iu}$ is a scaled version of  $\hat{\mathbf{w}}_{ib,iu}$ as in \eqref{thm2} and the scaler variable $\delta_{ib,u}$ depends on $\mathbf{w}_{ib,iu}$ itself. Instead of solving $\delta_{ib,u}$ from \eqref{eq: scaling equation} directly, we use the fact that the SINR constraint \eqref{eq: sinr constraint in JT-DL} in (JT-DL) must be satisfied with equality. This implies
\begin{align}\label{eq: sinr equality}
&\gamma_{iu}^{-1} \sum_{b \in \mathcal{C}_i} | \mathbf{h}^H_{ib,iu} \mathbf{w}_{ib,iu}|^2 = \sum_{b \in \mathcal{C}_i } \sum_{v \in \mathcal{U}_i \backslash u } | \mathbf{h}^H_{ib,iu} \mathbf{w}_{ib,iv} |^2 \nonumber \\
& \qquad + \sum_{\mathcal{C}_j \in \mathcal{N}_c \backslash \mathcal{C}_i } \sum_{b' \in \mathcal{C}_j } \sum_{v' \in \mathcal{U}_j} | \mathbf{h}^H_{jb',iu} \mathbf{w}_{jb',jv'} |^2 + \sigma^2.
\end{align}
Plugging \eqref{thm2} into \eqref{eq: sinr equality} gives
\begin{align}\label{eq: sinr equality-v2}
&\gamma_{iu}^{-1} \sum_{b \in \mathcal{C}_i} | \mathbf{h}^H_{ib,iu} \hat{\mathbf{w}}_{ib,iu}|^2 \delta_{ib,u} - \sum_{b \in \mathcal{C}_i } \sum_{v \in \mathcal{U}_i \backslash u } | \mathbf{h}^H_{ib,iu} \hat{\mathbf{w}}_{ib,iv} |^2 \delta_{ib,v} \nonumber \\
& \qquad  - \sum_{\mathcal{C}_j \in \mathcal{N}_c \backslash \mathcal{C}_i } \sum_{b' \in \mathcal{C}_j } \sum_{v' \in \mathcal{U}_j} | \mathbf{h}^H_{jb',iu} \mathbf{w}_{jb',jv'} |^2 \delta_{jb',v'} = \sigma^2,
\end{align}
which can be expressed in a matrix form as given by
\begin{equation}
\mathbf{F} \boldsymbol{\Delta} = \mathbf{1} \sigma^2
\end{equation}
where $\mathbf{F}$ is given in \eqref{fun:F}, $\boldsymbol{\Delta}$ is a vector collecting all $\delta_{ib,u}$ in \eqref{eq: sinr equality-v2}, and $\mathbf{1}$ is an all-one vector. Consequently, the scalar variables is obtained in~\eqref{fun:delta}. This completes the proof.

\section*{Acknowledgment}

This work was supported by the National Science and Technology Council, Taiwan, under Grant MOST 111-2221-E-007-076-MY3.

\bibliographystyle{IEEEtran}
\bibliography{ref}

\end{document}